\documentclass[12pt]{article}
\usepackage{graphicx}
\usepackage{wrapfig}

\newcommand\pubnumber{NuPhys2016-Liggins}
\newcommand\pubdate{\today}

\def\qmul{PPRC\\
Queen Mary University of London, London, UK}

\def\Title#1{\begin{center} {\Large #1 } \end{center}}
    \def\Author#1{\begin{center}{ \sc #1} \end{center}}
    \def\Address#1{\begin{center}{ \it #1} \end{center}}

    \newcommand\pubblock{\rightline{\begin{tabular}{l} \pubnumber\\
        \pubdate  \end{tabular}}}
\newenvironment{Abstract}{\begin{quotation}  }{\end{quotation}}
    \newenvironment{Presented}{\begin{quotation} \begin{center} 
    PRESENTED AT\end{center}\bigskip 
\begin{center}\begin{large}}{\end{large}\end{center} \end{quotation}}
    
\def\beq{\begin{equation}}
\def\eeq#1{\label{#1}\end{equation}}
\def\eeqn{\end{equation}}

\def\beqa{\begin{eqnarray}}
\def\eeqa#1{\label{#1}\end{eqnarray}}
\def\eeqan{\end{eqnarray}}

\let\bar=\overbar

\def\Dslash{\not{\hbox{\kern-4pt $D$}}}
\def\dslash{\not{\hbox{\kern-2pt $\del$}}}

\def\msb{{\bar{\ssstyle M \kern -1pt S}}}

\begin{document}
\begin{titlepage}
    \pubblock

    \vfill
    \Title{SNO+ Commissioning Status}
    \vfill
    \Author{ Billy Liggins on behalf of the SNO+ collabration}
    \Address{\qmul}
    \vfill
    \begin{Abstract}
        SNO+ inherits much of it's infrastructure from the Noble prize
        winning SNO detector. However due to the two orders of magnitude
        increase in light output in liquid scintillator many upgrades have been
        made. Details of these upgrades and their commissioning status are
        presented, along with an update on the current status of SNO+.
    \end{Abstract}
    \vfill
    \begin{Presented}
        NuPhys2016, Prospects in Neutrino Physics
        Barbican Centre, London, UK,  December 12--14, 2016
    \end{Presented}
    \vfill
\end{titlepage}
\def\thefootnote{\fnsymbol{footnote}}
\setcounter{footnote}{0}

\section{SNO+}

SNO+ is general purpose neutrino detector which will search for neutrinoless
double beta decay ($0\nu\beta\beta$) in 1.3\,T of $^{130}Te$ loaded in 780\,kT
of LABPPO \cite{whitePaper}.  SNO+ is situated at SNOLAB in Sudbury, Ontario,
Canada.  SNOLAB is 2\,km (6000 m.w.e) below ground in shaft 9 of the active
Creighton Nickel mine. At this depth cosmogenic backgrounds are reduced, with
the total muon flux less than $10^{-9}\,cm^{-2}s^{-1}$. SNO+ uses the same 6\,m
radius acrylic vessel (AV) as SNO and is observed by the same $\sim$~9300 8
inch Hamamatsu R1408 photomultiplier tubes providing a coverage of 54\%. The
detector is housed in a cavity filled with ultra pure water. While
commissioning the detector for $0\nu\beta\beta$ phase SNO+ will be sensitive to
invisible nucleon decay and solar axions during the initial water phase. During
the second phase, with a detector full of scintillator, SNO+ will search for
low energy solar neutrinos as well as Geo and reactor anti-neutrinos. SNO+
could also be sensitive to a galactic supernova event. A substantial amount of work
has been done to transform the detector from a heavy water based detector to a
liquid scintillator detector. 

\section{Detector and Processing Plant Upgrades}
\subsection{Hold Down Ropes}

As the LABPPO is less dense than water, the AV will experience a significant
buoyant force during scintillator fill. A hold down rope system was designed and installed in 2012 and
has been tested at various points throughout commissioning. Testing was achieved
by filling the cavity to a level above that of the AV fill level thus
manufacturing a buoyant force. The ropes system was found to work as expected
under an upward force of 1260\,kN.

\subsection{Scintillator Plant}

A new processing plant had been constructed to achieve the purity of
scintillator needed to ensure the low backgrounds required, $>10^{-17}
g/g_{LAB}$. Many tasks have been undertaken and completed; Helium leak checking,
cleaning and passivation, fire suppression system installation, pipe insulation
and water commissioning.  The plant is currently being commissioned with 40
tonnes of LABPPO. 

\subsection{Cover Gas}

The nitrogen cover gas that existed on SNO, has been upgraded to satisfy the
cleanliness requirements for low background. The cover gas seals the detector
from high levels of Radon gas in the mine air. The system also adjusts for
pressure differentials between the detector and the mine. The upgrades were
installed and commissioned in 2014.

\subsection{Universal Interface }
\label{sub:universal_interface_}

A new universal interface (UI) was designed and installed in late 2016. The UI
controls access to within the detector allowing the deployment of various
calibration sources, see \S \ref{sec::sources}. The UI also includes various
level sensors and veto PMTs. 

\section{Electronic Updates}

With the increased light output of liquid scintillator, the SNO+ electronics
have been upgraded to deal with the increase in current and trigger rate. New
XL3 readout cards can deal with the expected trigger rates, while also providing
ethernet communication with the front end boards. The MTC/A+ trigger cards can
handle the increase in PMT hits, as well as providing better baseline stability
and monitoring. The ability to introduce reprogrammable trigger logic has also
been added. The CAEN v1720 digitizer aids in instrumental background reduction
through outputting the triggered waveforms on each event. An additional trigger
utility board, TUBii, adds an extensive suite of tools tying many parts of the
experiment together. TUBii is built around a MicroZed development board
containing a Zynq chip running a FPGA along side a Linux processing system.
Features include synchronisation of calibration systems with detector readout,
extra trigger ports and detector wide timing verification. TUBii also introduces
on the fly programmable trigger logic. The full electronics system (with
upgrades) has been testing and commissioned in all phases to date. 

\section{Data Acquisition Upgrades}

The data acquisition system (DAQ) has undergone a complete overhaul, which saw
improvements in decoupling data flow from detector controls (ORCA). Taking this
modular approach provides improved detector control and stability. Stress tests
have taken place on various occasions throughout air and partial water fill
phases. The detector control GUI provides both user and expert operation modes,
giving increased confidence in detector output. Online monitoring tools have
been developed capable of displaying data at 20\,kHz. These tools offer the
possibility of monitoring the detector from anywhere in the world, added to a
potential remote shifters tool box. An extensive alarm sever with accompanying
GUI, which monitors both short and long time changes has been implemented.
Improved detector readout during ramping the detector to high voltage has been
implemented, providing a responsive GUI allowing controllers to check for
abnormalities channel to channel. The detector state is committed to a database
on a run by run bases, allowing for individual run reproducibility during
offline analyses. Grid based processing and storage channels have been updated
accounting for the increased trigger rate. The full DAQ has been stress tested
in various ``mock data challenges'' which have helped in finding potential
bottle necks and speed ups. 

\section{Calibration} 
\subsection{In situ sources}

The Embedded LED/Laser Light Injection Entity (ELLIE) is a new multi purpose
optical calibration system. It consists of three sub systems, AMELLIE, SMELLIE
and TELLIE.  AMELLIE will measure attenuation in scintillator, thus monitoring
scintillator quality, and is due to come online for scintillator phase. SMELLIE
uses a super continuum laser to measure the scattering length of the detecting
medium. SMELLIE will run across all phases of data taking and at the time of
writing is currently being commissioned. TELLIE uses high frequency LEDs to
measure the timing response of the PMTs and was fully commissioned in the
spring of 2017. For more details on the in suit calibration in SNO+ see
\cite{ELLIE}.

\subsection{Deployed sources}
\label{sec::sources}
Substantial effort has gone into updating the deployed calibration sources from
SNO ready for the SNO+ scintillator phase. Due to the increased cleanliness
requirements new deployment mechanisms have been developed, machined and
shipped to SNOLAB. A new ``laserball'' reducing the self shadowing angle from
30$^\circ$ to 7$^\circ$ has also been designed and constructed.

\section{Current Status}

SNO+ is currently in operational mode currently undergoing water calibration.
The detector electronics were switched on in early 2017 and have been are
functioning and stable. Water fill is due to finish in late spring 2017, however
the water level has been above the PMTs for some time allowing for calibration
to take place. Both in situ and deployed source calibration is on going. 

\section{Outlook}

With calibration complete SNO+ will search for invisible nucleon decay as well
as characterising the state of the detector. Scintillator fill is scheduled to
commence in August 2017 and to be completed by the end of the year. The
Scintillator phase will build upon the work done in the previous phase and
assess the background levels present in the detector, optical studies will also
take place. Reactor and Geo anti-neutrino studies will be undertaken and
background level permitting low energy Solar signals will be measured. With the
background studies complete, loading with $^{130}Te$ will begin, thus commencing
the $0\nu\beta\beta$ phase. This is scheduled for early 2018.




\end{document}